**Phase stability and mechanical response of Ag-interlayered Al/Cu resistance spot-welded joints**

Shuang Lin[a], Kyubok Lee[b], Jiahui Ye [b], Ho Kwon[c], Shun-Li Shang[a,*], Allison M. Beese[a,d], Xun Liu[c], Jingjing Li[b,*] and Zi-Kui Liu[a]

[a] Department of Materials Science and Engineering, The Pennsylvania State University, University Park, PA, 16802, USA

[b] Department of Industrial and Manufacturing Engineering, The Pennsylvania State University, University Park, PA 16802, USA

[c] Welding Engineering, Department of Materials Science & Engineering, The Ohio State University, Columbus, OH 43202, USA

[d] Department of Mechanical Engineering, The Pennsylvania State University, University Park, PA 16802, USA

*Corresponding authors. E-mail addresses: sus26@psu.edu (Shun-Li Shang); jul572@psu.edu (Jingjing Li)

**Keywords:** Resistance spot welding; Al/Cu joints with Ag interlayer; Intermetallic compound suppression; Mechanical properties; First-principles calculations; CALPHAD

**Abstract**

Dissimilar Al/Cu joints are essential to battery-pack assemblies; however, their mechanical strength is limited by brittle Al-Cu intermetallic compounds (IMCs) such as $Al_2Cu$ and $Al_4Cu_9$. Interlayer strategies to suppress these phases remain largely empirical, lacking a predictive framework linking interlayer chemistry to the phases that form and to their intrinsic mechanical character. Here an Ag interlayer is introduced and combines computational thermodynamics, first-principles calculations, microstructural characterization, and mechanical testing into a single self-consistent description of the joint. CALculation of PHAse Diagrams (CALPHAD) equilibrium and Scheil simulations predict the solidification path of the Al-rich Al-Ag fusion zone and explain why Cu incorporation is limited when Ag is present; energy-dispersive X-ray spectroscopy (EDS) and electron backscatter diffraction (EBSD) confirm an FCC Al-Ag solid solution as the dominant constituent. First-principles phonon calculations within the quasiharmonic approximation yield finite-temperature entropy and Gibbs energy, benchmarked against CALPHAD, while elastic constants assess ductility via the Pugh criterion (i.e., the bulk/shear (B/G) modulus ratio). All Al-Ag phases, including the observed solid solution, exceed the Pugh threshold of 1.75, whereas the targeted Al-Cu IMCs do not, giving a mechanistic basis for the interlayer's effectiveness. This microstructural change translates into improved performance: nominal strength rises from 47.9 to 67.4 MPa. Nanoindentation gives a fusion-zone reduced modulus of 82.8 GPa (Young's modulus 82.0 GPa), versus a calculated 0 K Voigt-Reuss-Hill value of 71.4 GPa. The present work establishes a transferable CALPHAD, first-principles, and experiment workflow for rational interlayer selection in dissimilar-metal joining.

## 1. Introduction

The joining of aluminum to copper has become a recurring requirement in the assembly of electric-vehicle battery packs, where aluminum conductors are welded to copper tabs and busbars to carry high currents at lower weight and cost [1,2]. A single pack contains hundreds to thousands of such interconnects, and each one adds to the electrical loss, the heat generation, and the mechanical durability of the system [1]. Copper carries about 1.6 times as much current as aluminum for a given cross-section, near 58 against 37 MS $m^{-1}$, and it removes heat more effectively, near 401 against 237 W $m^{-1}$ $K^{-1}$, but it is about 3.3 times denser, 8.96 against 2.70 g $cm^{-3}$, and more expensive by weight [3,4]. Aluminum is therefore used in place of copper where mass and cost matter, and the two metals are joined rather than substituted. The performance of the pack then depends on the quality of the dissimilar joint between them.

The difficulty of this joint is metallurgical. Aluminum and copper have limited mutual solid solubility, and when they meet in the liquid state, they react to form a sequence of intermetallic compounds (IMCs), most often $\theta$-$Al_2Cu$, $\eta$-AlCu, and $\gamma$-$Al_4Cu_9$ [5]. These phases have low-symmetry crystal structures with few active slip systems, and they are correspondingly hard and brittle, with indentation hardness above 10 GPa [6]. They also conduct poorly, with the Al-Cu IMCs several times more resistive than the parent metals [7]. Once a continuous intermetallic layer grows beyond about 2 to 2.5 μm, the joint tends to fail along the reaction layer and both its strength and its conductivity fall sharply [8]. Controlling the amount and type of IMCs that forms is therefore the central problem in Al/Cu joining.

Two broad routes have been used to limit intermetallic growth. The first shortens the time spent in the reactive temperature range, either through low-heat solid-state processes such as friction stir welding, ultrasonic welding, and magnetic pulse welding [9,10], or through close control of heat input in fusion processes [11]. The second inserts a third material as an interlayer that keeps aluminum and copper apart and redirects the interfacial reaction toward less harmful products. Nickel, zinc, tin, and graphene interlayers have each been reported to reduce Al/Cu intermetallic formation and to raise joint strength [12–15]. Silver is a reasonable choice for this role. It is highly conductive, more so than copper, and therefore does not introduce an intrinsic bulk-conductivity disadvantage, and it separates the direct Al/Cu reaction into the Al-Ag and Ag-Cu subsystems, which offer broader terminal solid-solution fields and avoid the sequence of brittle compounds seen in Al/Cu over the relevant composition range. Silver interlayers have already been reported to reduce Al/Cu intermetallic formation and to raise joint strength in laser welding [16] and in friction stir welding [17].

Resistance spot welding is well suited to battery assembly because it is fast, easily automated, and inexpensive, but its high heat input makes intermetallic control demanding [18]. Sun et al. [19] used a CALPHAD-based composition-pathway tool to design a route from aluminum to copper through a silver interlayer that avoids the brittle Al-Cu phases, and demonstrated the feasibility of this joint by resistance spot welding with microstructural characterization. The present study is complementary and addresses a different question: the finite-temperature thermodynamic properties and elastic properties of Al-Ag phases that form in the fusion zone, and how these properties relate to the local response measured by nanoindentation and to the strength of the joint. A phase-diagram argument identifies which phases are avoided, but it does not give the finite-temperature stability of the phases that take their place, nor their relative tendency toward ductile or brittle elastic behavior. These properties follow from the electronic structure and lattice dynamics of the phases.

The present work addresses this point by examining the Al-Ag phases of a silver-interlayer Al/Cu spot weld at the atomic scale, cf., the predicted Al-Cu, Al-Ag, and Cu-Ag phase diagrams in Figure 1 in terms of the database developed by Witusiewicz et al. [20]. The preceding study [19] established the CALPHAD-guided silver composition-pathway design; the resistance spot welds, the microstructural characterization, and the mechanical measurements reported here were produced and obtained specifically for the present study, and no experimental data are reproduced from [19]. The finite-temperature thermodynamics properties and the elastic analysis are its principal new contribution. First-principles calculations using special quasirandom structures (SQS) [21] give the formation energies and ground-state elastic constants of the candidate Al-Ag phases, the DFT-based quasiharmonic approach in terms of phonon calculations extends their entropy and Gibbs energy to finite temperature [22], and the Pugh ratio of bulk to shear modulus (B/G) is used as an empirical indicator of ductile versus brittle elastic behavior [23]. The predictions are compared with CALPHAD data and are tested against energy-dispersive X-ray spectroscopy, electron backscatter diffraction, tensile testing, and nanoindentation of joints welded with and without a silver interlayer. The purpose of the present work is to determine whether the calculated thermodynamic and elastic properties of the Al-Ag phases are consistent with the experimentally observed fusion-zone structure and joint response.

## 2. Computational methods

### 2.1 Equilibrium calculations and Scheil-Gulliver simulations by CALPHAD

In the present work, equilibrium and Scheil-Gulliver (hereafter referred to as Scheil) calculations were used to predict phase formation in the Al-Ag system. The CALPHAD method [24,25] models the Gibbs energy of phases as functions of temperature, pressure, and composition; the calculations here followed the thermodynamic description and computational implementation of the preceding study [19]. Equilibrium calculations predict phases under slow cooling, while Scheil simulations account for rapid cooling by assuming local equilibrium at the solid/liquid interface, complete mixing in the liquid, and no diffusion in the solid [26]. Scheil simulations capture the evolving liquid composition during solidification, enabling the prediction of non-equilibrium phases, particularly relevant to welding and additive manufacturing. In addition, the hybrid Scheil-equilibrium method was used to couple Scheil solidification simulations with equilibrium phase calculations on the segregated compositions [27]. This enables prediction of phase formation in the last-solidifying regions and provides a more sensitive indicator of solidification-related defects. The Al-Ag system is the focus of these analyses because the silver interlayer makes the fusion zone essentially an Al-Ag mixture, while the Ag-Cu system forms no IMCs, so the Al-Ag phases are relevant to the local mechanical response of the characterized fusion zone.

### 2.2 First-principles quasiharmonic phonon approach

Density functional theory (DFT) based first-principles calculations provide quantities linked to electronic structure. Helmholtz energy ($F$) of a given configuration as a function of volume ($V$) and temperature ($T$) within the quasiharmonic approach (QHA) is represented by [22,28],

$$F(V, T) = E_0(V) + F_{vib}(V, T) + F_{el}(V, T) \quad (1)$$

Here $E_0(V)$ is the static energy at 0 K, excluding the zero-point vibrational energy, obtained directly from DFT; $F_{vib}(V, T)$ is the vibrational contribution to the Helmholtz energy; and $F_{el}(V, T)$ is the thermal electronic contribution, obtained by integrating over the electronic density of states (DOS) with the Fermi-Dirac distribution. The electronic term is included here because the Al-Ag phases are metallic. It should be noted that, under the zero external pressure conditions considered in the present work, the Helmholtz energy is identical to the Gibbs energy.

The present work adopts the four-parameter Birch-Murnaghan equation of state (EOS) in energy form [22] to model the energy-volume (E-V) relation from DFT-calculated data points,

$$E(V) = a + b\,V^{-2/3} + c\,V^{-4/3} + d\,V^{-2} \tag{2}$$

where a, b, c, and d are fitting parameters. From this fit, four equilibrium properties at zero external pressure are obtained: the equilibrium volume $V_0$, the equilibrium energy $E_0$, the bulk modulus $B_0$, and its pressure derivative $B_0'$. The vibrational Helmholtz energy $F_{vib}(V, T)$ is obtained from the DFT phonon density of states [22,28],

$$F_{vib}(V,T) = k_B T \int \ln\{2\sinh[\hbar\omega/2k_B T]\}\; g(\omega)\, d\omega \tag{3}$$

where $k_B$ is the Boltzmann constant, ħ is the reduced Planck constant, ω is the phonon frequency, and g(ω) is the phonon density of states.

The single-crystal elastic stiffness constants ($C_{ij}$) were obtained using the efficient stress-strain method suggested by Shang et al. [29,30] . A set of small, symmetry-preserving strains $\varepsilon$ is applied to the lattice, the resulting stresses $\sigma$ are computed for each deformed structure, and the stiffness matrix $C$ follows from Hooke's law: $\sigma = C\,\varepsilon$, hence

$$C = \sigma \cdot \varepsilon^{-1} \tag{4}$$

In the present work, six independent strain patterns of magnitude ±0.01 were applied to each structure and the elastic constants obtained by least-squares fitting of the resulting stresses. For the SQS supercells with monoclinic symmetry, the stiffness matrix was symmetrized to the parent cubic or hexagonal symmetry of the underlying FCC, BCC, or HCP lattice before the Voigt-Reuss-Hill averages were evaluated. The elastic constants were evaluated at the relaxed 0 K equilibrium volume, so the reported moduli are ground-state values and are not corrected for thermal softening, which for Al-rich phases lowers the moduli by roughly 5 to 8% between 0 K and room temperature.

2.3 Details of first-principles and phonon calculations

All DFT-based calculations were performed with the Vienna Ab initio Simulation Package (VASP) [31] using the projector augmented wave (PAW) method [32] and the generalized gradient approximation developed by Perdew, Burke, and Ernzerhof (PBE) for exchange and correlation [33]. The electronic self-consistency criterion was $10^{-6}$ eV for the total energy of the cell and the plane-wave cutoff was 350 eV, about 1.4 times the larger of the two recommended PAW cutoffs (249.8 eV for Ag and 240.3 eV for Al). The disordered FCC, BCC, and HCP solid solutions were represented by 16-atom special quasirandom structures (SQS) [21], with the BCC phase treated as a disordered A2 solution rather than as an ordered B2 compound, and the A13 phase by its cubic unit cell with 20 atoms. The SQS and phonon supercells were constructed following the same computational framework as the preceding study [19]. For the static energy-volume and elastic-constant calculations, Monkhorst-Pack k-point meshes [34] of 15×15×15 (FCC), 14×14×8 (HCP), and 12×12×12 (BCC and A13) were used, and all converged to within 1 meV/atom. Phonon calculations used the supercell method and the aforementioned supercells with a 6×6×6 k-point mesh and were carried out with the YPHON package [35] coupled to VASP.

## 3. Experimental methods

### 3.1 Resistance spot welding (RSW)

The welding was performed using a resistance spot welding (RSW) machine equipped with a medium-frequency direct-current (MFDC) inverter (T.J. Snow Co, Inc). The configuration and dimensions of the weld stacks are shown in Figure 2a and were designed to produce two dissimilar weld nuggets, at the top and at the bottom, under the same programmed welding schedule. Two Ag foils were stacked together and placed at the top interface, between the Cu tab and the Al foil, to investigate the effects of Ag on solidification behavior. The thicknesses of the Cu tab, the Ag foils, the Al foils, and the thick Al sheet are given in Figure 2a, together with the electrode material and face diameter. In this configuration, the top weld nugget reflects the chemical reactions among Cu, Ag, and Al, while the bottom weld nugget reflects the interactions between Al and Cu. Additionally, an extra Al foil was placed above and below the thick Al sheet to increase electrical resistance, thereby concentrating heat generation at the weld interface.

A schematic of the welding procedure is shown in Figure 2b. The sample was subjected to a preset weld force of 1.8 kN, calibrated to retain the molten material without expulsion. Two preheating pulses were applied at a current of 2 kA for 100 ms to minimize interlayer gaps and remove surface asperities, thereby reducing contact resistance and improving process reproducibility. Welding was then performed at 11 kA for 100 ms, followed by a holding time of 700 ms. The welding process parameters and configuration were carefully selected to suppress the formation of an excessively large fusion zone while achieving localized melting at the Cu interface, considering the significant difference in melting points between Al (660 °C) and Cu (1085 °C). Because the Al/Ag/Cu and Al/Cu stacks differ by the thickness of the two Ag foils, their total electrical resistance is not identical, so the same programmed schedule does not deliver exactly the same heat input to the two conditions.

### 3.2 Material characterization

SEM, EBSD, and EDS analyses were performed to characterize crystallographic orientation, grain structure, and chemical composition of the welded samples. The specimens were prepared by mechanical grinding and polishing to achieve a smooth surface suitable for EBSD and EDS analysis. EBSD measurements were conducted using an FEI Apreo 2 system operated at an accelerating voltage of 15 kV, with a step size of 1 µm for the broader area (275 × 275 µm) and 0.15 µm for the smaller region (80 × 30 µm) to ensure high-resolution mapping, particularly around the weld interface. The EBSD detector was positioned at a 70° tilt relative to the sample surface to optimize backscatter diffraction collection. Data acquisition and post-processing were carried out using Aztec software (Oxford Instruments), with noise reduction techniques applied to enhance the quality of the EBSD orientation maps and to improve the accuracy of the EDS chemical composition analysis.

### 3.3 Mechanical testing

A quasi-static tensile test was used to evaluate the strength of the resistance spot welded joints. Tests were carried out on both Al/Cu and Al/Ag/Cu samples so that the two joint types could be compared. The specimen geometry is shown in Figure 2c. Loading was applied with an MTS 800 test system at a constant crosshead speed of 2 mm/min, with the aluminum stack clamped on one side and a single copper tab on the other, so that the load passed through one weld nugget. Because the specimen is a spot-welded lap joint rather than a uniform gauge section, the reported quantities are nominal. The nominal stress is the peak load divided by the weld area A0, taken as the projected area of the weld and measured individually on each fractured specimen, and the abscissa is the crosshead displacement normalized by the initial specimen length $l_0$. These curves describe the response of the joint and are not the true stress-strain behavior of a homogeneous material. Each condition was tested three times, and the fracture mode of every specimen was recorded after testing.

3.4 Nanoindentation

Nanoindentation was conducted to assess the localized mechanical properties of the welded samples. The test was performed using a Bruker Hysitron TI-980 system with a diamond Berkovich indenter to measure the indentation modulus by the Oliver-Pharr method [36]. The indentation load was set to 2 mN, with a dwell time of 0.3 seconds at the peak load to ensure accurate measurement. Indentation was performed at various locations across the weld fusion zone to provide a comprehensive profile of mechanical property variation. Eight indents were performed, and both the individual values and their mean values are reported. The Oliver-Pharr analysis yields the reduced modulus; the corresponding specimen Young's modulus $E$ was obtained by removing the compliance of the diamond indenter using the calculated Poisson's ratio of the FCC Al-Ag solid solution given in Table 1.

**4. Results and discussion**

4.1 Phase stability and thermodynamic properties of the Al-Ag system

Figure 3 shows the formation enthalpy ($\Delta H_0$) values for Al-Ag compounds as predicted by the present DFT calculations at T = 0 K and P = 0 GPa. The phases on the convex hull include Ag (FCC), $Ag_3Al$ (HCP), $Ag_{13}Al_7$ (A13), AgAl [BCC (A2, disordered)], and pure Al (FCC). Detailed $\Delta H_0$ values are summarized in Table 1. Overall, the predicted formation enthalpies agree, to within about 1 kJ per mole of atoms, with the experimental data reported by Wittig and Schilling [37] and by higher-temperature calorimetry [38], although the experimental values themselves show some scatter. The DFT predictions show particularly high accuracy for dilute FCC Ag, capturing the stability trends well. Dilute Al alloys containing up to 10% Ag have small positive formation enthalpies at 0 K and therefore lie above the convex hull. This is not inconsistent with the measured terminal solubility of Ag in FCC Al, because the configurational entropy of mixing stabilizes these compositions at the temperatures at which that solubility is observed. For the $Ag_3Al$ compound, the DFT calculations predict a higher stability compared to experimental observation. This discrepancy, up to 0.8 kJ $mol^{-1}$ per mole of atoms, is within the overall scatter of the comparison quoted above and is consistent with the experimental measurements having been conducted at elevated temperatures of 743 K and 955 K, respectively, rather than at 0 K.

Two thermodynamic properties — entropy and Gibbs energy — at finite temperatures were evaluated using DFT-based phonon calculations and compared with CALPHAD-based results. The most stable phases, including AgAl (BCC), $Ag_3Al$ (HCP), and $Ag_{13}Al_7$ (A13), were identified through these calculations. Ground-state elastic constants and polycrystalline elastic moduli were calculated separately from the 0 K DFT results. In addition, the ideal configurational entropy ($S_{conf}$) was added to the phonon-derived entropy to account for the chemical disorder that the SQS represents but that the harmonic phonon spectrum of a single fixed configuration does not contain. The correction is applied as a temperature-independent term, which is the high-temperature random-mixing limit; it is therefore not valid as T approaches 0 K, where the third law of thermodynamics requires the configurational contribution to vanish, and the corrected curves in Figure 4, Figure 5, and Figure 6 should be read only above about 300 K. The ideal configurational entropy was calculated using the following relation:

$$S_{conf} = -R \sum_s n_s \sum_i y_i^s \ln y_i^s \quad (5)$$

where $R$ is the gas constant, $y_i$ is the site fraction of species i on sublattice s, $n_s$ is the number of sites on that sublattice per mole of atoms, and the summation runs over the partially occupied Wyckoff sites. For the disordered BCC and HCP solutions a single mixing sublattice was used, giving the random-mixing values at the relevant compositions. The calculated configurational entropy values for AgAl (BCC), $Ag_3Al$ (HCP), and $Ag_{13}Al_7$ (A13) are 5.76 J/(mol·K), 4.68 J/(mol·K), and 1.25 J/(mol·K), respectively.

Figure 4 compares thermodynamic properties of the AgAl (BCC) phase predicted by the quasiharmonic approach, phonon calculations incorporating ideal configurational entropy corrections, and the CALPHAD data. The phonon-only predictions underestimate the entropy by about 4 J $K^{-1}$ $mol^{-1}$ relative to the CALPHAD data, particularly at higher temperatures. When the configurational entropy correction (Phonon+$S_{conf}$) is included, the predicted entropy increases and closely matches the CALPHAD data, as highlighted in the zoomed-in region. Because the phonon-only calculation underestimates the entropy, its Gibbs energy lies above the CALPHAD values; including the configurational entropy lowers the Gibbs energy and brings it into closer agreement across the temperature range. In the BCC AgAl phase, substantial partial occupancy on Wyckoff sites leads to a larger configurational entropy contribution, making the correction essential for accurate thermodynamic predictions.

For the HCP phase ($Ag_3Al$), Figure 5 shows that the difference between phonon-only predictions and CALPHAD results is comparatively smaller within 0.2 J $K^{-1}$ $mol^{-1}$. Incorporating configurational entropy slightly worsens the agreement by further increasing the predicted entropy beyond CALPHAD values. This suggests that the ideal configurational-entropy correction overestimates the CALPHAD entropy for the HCP phase, so the actual degree of configurational disorder is lower than the fully random limit assumed in the ideal model.

For the A13 phase ($Ag_{13}Al_7$), Figure 6 shows that both the phonon-only and the corrected predictions exhibit very close agreement with the CALPHAD data throughout the entire temperature range. The configurational entropy correction (Phonon+$S_{conf}$) provides only a minor adjustment (1.25 J $K^{-1}$ $mol^{-1}$), slightly improving the match. The Gibbs energy trends similarly show that the phonon-only predictions are marginally more negative, and the inclusion of configurational entropy leads to an even closer agreement. Despite its structural complexity, the A13 phase has nearly full, ordered occupancy of its Wyckoff sites, which gives a small configurational entropy. The relatively small magnitude of correction indicates that atomic disorder plays less role in determining thermodynamic properties of the A13 phase.

4.2 Phase and microstructure of the Al-Ag-Cu fusion zone

Figure 7 shows the atomic distributions of Al, Ag, and Cu across the Al/Ag/Cu joint measured by EDS line scanning. The three scans (Scan 1, Scan 2, and Scan 3) give broadly similar Al-rich profiles, although some local compositional variation remains. The average contents are 84.4 at.% Al (with standard deviation SD of 6.5), 14.7 at.% Ag (SD 5.9), and 0.8 at.% Cu (SD 2.2). This distribution follows from the melting sequence during resistance spot welding. Aluminum melts first at 660 °C and forms a large molten pool; silver, which melts at 961 °C, then melts near the weld center and mixes with the aluminum to give a broadly Al-rich Al-Ag composition, while toward the edges the silver layer stays largely intact, as seen in Scan 3. Copper, with the highest melting point at 1085 °C, melts only locally at the interface, so its participation in the pool is limited and its content remains low throughout the joint. The low average Cu content is consistent with limited Cu incorporation into the characterized Al-rich fusion zone. The reported standard deviations describe spatial composition variation along the line scans rather than specimen-to-specimen variability.

The CALPHAD method was used to predict the phases in the fusion zone from the measured compositions. Because the Cu content of the fusion zone was below 1 at.%, the phase analysis was approximated using the binary Al-Ag subsystem. Considering the compositional variation (standard deviation) observed in the line scan, five representative compositions ($Al_{60}Ag_{40}$, $Al_{65}Ag_{35}$, $Al_{70}Ag_{30}$, $Al_{75}Ag_{25}$, and $Al_{80}Ag_{20}$) were evaluated to examine the composition-dependent solidification trend in the Al-rich portion of the binary system, using both equilibrium and Scheil simulations. The mean EDS composition was slightly more Al-rich than this calculated grid, so the CALPHAD results are interpreted as a qualitative composition trend rather than an exact phase-fraction prediction at the measured mean composition. Under equilibrium conditions, the calculations indicate that a fully FCC microstructure is reached on cooling below approximately 720 K, as shown in Figure 8. The Scheil simulations reveal a change in the solidification sequence between the $Al_{70}Ag_{30}$ and $Al_{65}Ag_{35}$ compositions: at lower Al concentrations, solidification initiates with the formation of an HCP phase, which subsequently transforms into the FCC phase (Figure 9a). In the hybrid-Scheil simulations, Ag is allowed to reach equilibrium by back-diffusion in the solid, while Al, as the slower-diffusing substitutional species, follows the Scheil assumption of no solid-state diffusion. It should be noted that the Scheil construction assumes local equilibrium at the solid-liquid interface, an assumption that becomes questionable at the cooling rates of order $10^3$ to $10^4$ K $s^{-1}$ expected in resistance spot welding, so the calculated non-equilibrium phase fractions are best regarded as bounding estimates. As shown in Figure 9b, with increasing Al concentration, the FCC phase becomes increasingly dominant. For example of the $Al_{80}Ag_{20}$ composition, approximately 91% FCC phase and 9% HCP phase are predicted. The EBSD measurements described below resolved only FCC in the fusion zone, so the predicted minority HCP fraction was either not formed under the actual weld cooling conditions or was present at a scale and fraction below the detection limit of the mapping conditions used.

The phases and microstructure of the Al/Ag/Cu joint were examined by EBSD. As shown in Figure 10b and Figure 10c, EBSD identified an FCC crystal structure in the fusion zone, and the EDS composition indicates that this is an Al-rich Al-Ag solid solution. This is consistent with the CALPHAD and first-principles predictions: during welding, silver and aluminum interdiffuse and react to form a thermodynamically stable FCC solid solution. Region 2 in Figure 10d and Figure 10e was scanned at higher resolution to resolve the interface, where fine grains are observed. These may be associated with localized melting and rapid solidification of the copper surface during welding.

4.3 Mechanical properties of Al-Ag-Cu fusion zone

The bulk modulus (B) and shear modulus (G) represent the resistance to volume and shape deformation, respectively. Their ratio (B/G) serves as an indicator of material ductility, as proposed by Pugh [23]. B/G ratios for the HCP, A13, BCC, and FCC phases are shown in Figure 11, and the detailed Voigt-Reuss-Hill (VRH) [39] averaged values are listed in Table 1. The dashed line at B/G = 1.75 is the empirical Pugh criterion, a rough boundary between ductile (B/G > 1.75) and brittle behavior [23]. For the phases studied here, the data points lie above the Pugh line, indicating a tendency toward ductile rather than brittle elastic behavior. The FCC Al-Ag solid solution actually observed in the fusion zone has B/G = 3.2, well above the threshold. The very high B/G value of the $Ag_3Al_2$ (A13) composition reflects a shear modulus of only 9.1 GPa, which is anomalously low for a metallic phase; this structure lies off the convex hull and its elastic response is better regarded as indicating incipient mechanical instability than as exceptional ductility. The Pugh ratio is used here only as an empirical elastic indicator evaluated at 0 K and is not a direct prediction of the fracture ductility of the welded joint [40].

Tensile testing was used to compare the two joint types. Figure 12 shows the nominal joint stress against normalized crosshead displacement for Al/Cu joints welded with and without a silver interlayer. Because the load in a spot weld is carried across a small, bonded area under a combined peel and shear state, the values reported here are nominal joint strengths, defined as the peak load divided by the weld area, rather than a true uniaxial ultimate tensile strength. On this basis the joints with the silver interlayer reached an average nominal strength of 67.4 MPa over three tests, against 47.9 MPa for the joints without it. The mean nominal joint strength was higher for the silver-interlayer condition than for the condition without silver. Because only three tests were carried out for each condition, this comparison is descriptive and no formal statistical significance is claimed. This higher strength is associated with the presence of the silver interlayer and the formation of an Al-rich FCC Al-Ag fusion zone, and the low Cu content in the characterized fusion zone is consistent with reduced direct Al-Cu interaction. Because the interfacial reaction layer was not characterized in cross-section, the specific strengthening mechanism cannot be established from the present data. Because the two stacks differ in total thickness and therefore in heat input, part of the difference in nominal strength may also reflect a difference in nugget size rather than in interfacial metallurgy. The displacement at peak load is of a similar order for the two joints, so the effect of the interlayer is seen in the peak load rather than in the overall compliance of the response.

Nanoindentation was used to probe the local stiffness of the fusion zone. Figure 13 shows the load-displacement curves for the silver-interlayer joint, from which the average reduced modulus of the fusion zone is 82.8 GPa. Removing the compliance of the diamond indenter and using the calculated Poisson's ratio of 0.36 gives a specimen Young's modulus of 82.0 GPa. For comparison, the Voigt-Reuss-Hill Young's modulus of the FCC Al-rich Al-Ag solid solution ($Al_{75}Ag_{25}$ SQS) is about 71 GPa, obtained from the calculated bulk and shear moduli listed in Table 1. The measured modulus is therefore about 15% higher than the calculated value. Several effects contribute to this difference and act in the same direction: the calculated value is a 0 K ground-state result whereas the measurement is at room temperature, which lowers rather than raises the true modulus; pile-up around the indent in a soft FCC solid solution causes the Oliver-Pharr method to underestimate the contact area and so to overestimate the modulus, typically by 10 to 30% in such materials; and the indents sample a compositionally variable region and are influenced by the surrounding material. The comparison should accordingly be read as agreement in magnitude rather than as quantitative validation. This provides a comparison of the experimental and calculated stiffness scales; the phase itself is identified from the EDS and EBSD results.

## 5. Conclusions

In this study, the effects of introducing an Ag interlayer on the thermodynamics, microstructure, and mechanical performance of Al/Cu spot-welded joints were systematically investigated. First-principles calculations and CALPHAD modeling were employed to predict phase stability, thermodynamic properties, and elastic behavior of Al-Ag compounds. The DFT-calculated thermodynamic properties (enthalpy, entropy, and Gibbs energy) showed reasonable agreement with the CALPHAD description over the investigated temperature range, supporting the stability of the key phases $Ag_3Al$ (HCP), $Ag_{13}Al_7$ (A13), and AgAl (BCC). The configurational entropy correction substantially improved the agreement for BCC AgAl and produced only a minor change for the A13 phase, whereas the ideal random-mixing approximation overestimated the entropy of HCP $Ag_3Al$. Experimental characterizations via EDS and EBSD were consistent with the computational predictions, indicating that the fusion zone consisted mainly of a predominantly Al-rich FCC Al-Ag solid solution with some local compositional variation. CALPHAD-based equilibrium and Scheil simulations demonstrated that with increasing Al concentration, the FCC phase becomes dominant in the fusion zone. These results are consistent with reduced direct Al-Cu interaction in the presence of silver and with the formation of an Al-rich FCC Al-Ag solid solution whose calculated elastic indicators favor ductile rather than brittle behavior. Mechanical testing showed a higher joint strength with the Ag interlayer. The nominal strength of the spot-welded joint increased in the tested specimens from a mean of 47.9 MPa to 67.4 MPa with the silver interlayer. Nanoindentation gave an average reduced modulus of 82.8 GPa in the fusion zone, corresponding to a specimen Young's modulus of 82.0 GPa, about 15% higher than the calculated 0 K value of 71.4 GPa for the Al-Ag solid solution; indentation pile-up is the most likely origin of the remaining difference. Elastic analysis gave B/G ratios above the empirical Pugh threshold for the Al-Ag phases, indicating a tendency toward ductile rather than brittle elastic behavior. These results support, but do not directly prove that reduced direct Al-Cu interaction contributes to the higher joint load; the interfacial reaction layer and the fracture path were not characterized in this work.

**Acknowledgements**

This work was funded by the U.S. National Science Foundation (NSF) via Award No. CMMI-2226976. First-principles calculations were performed through allocation DMR140063 from the Advanced Cyberinfrastructure Coordination Ecosystem: Services & Support (ACCESS) program, which is supported by U.S. National Science Foundation grants #2138259, #2138286, #2138307, #2137603, and #2138296.

## Figures and Figure Caption

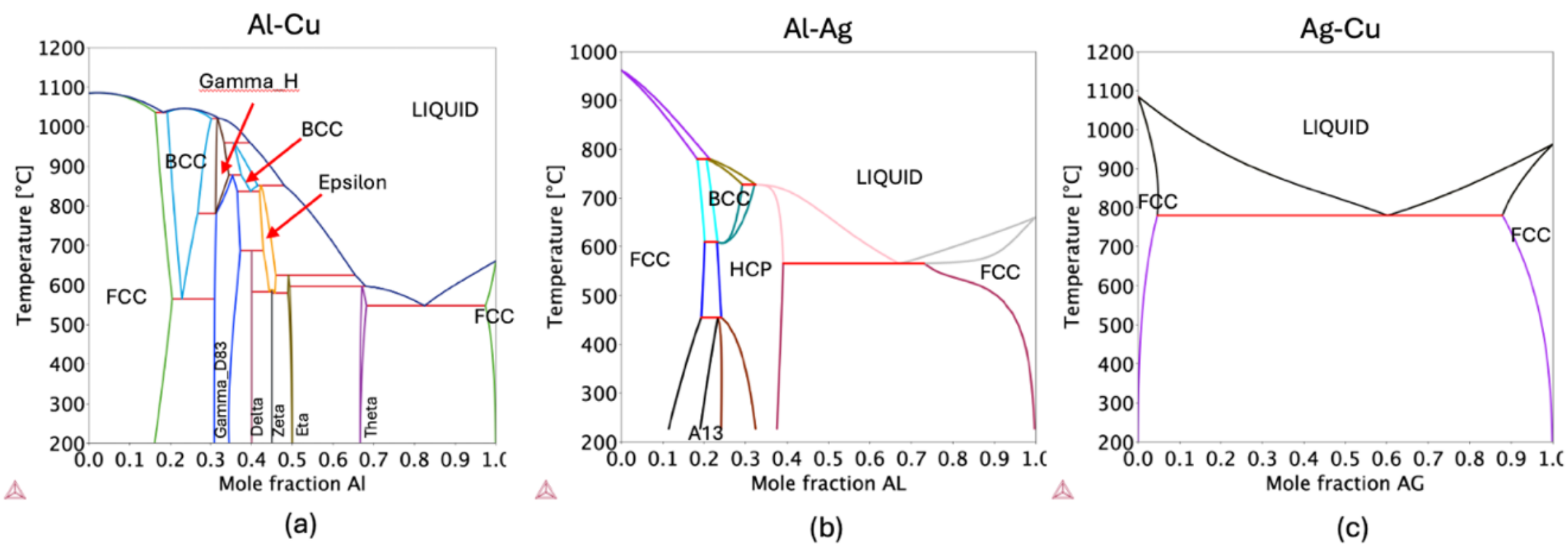

Figure 1. Predicted binary phase diagrams of (a) Al-Cu, (b) Al-Ag, and (c) Ag-Cu systems in terms of database by Witusiewicz et al. [20]. In the Al-Cu system (a), seven intermetallic compounds (IMCs) are observed, in addition to the liquid, FCC, and BCC phases. In contrast, the Al-Ag phase diagram (b) reveals mutual solubility between Al and Ag, with stable FCC, BCC, and HCP solid-solution phases and an A13 intermediate phase at lower temperatures. The Ag-Cu system (c) shows the formation of only FCC phases between Ag and Cu without IMCs.

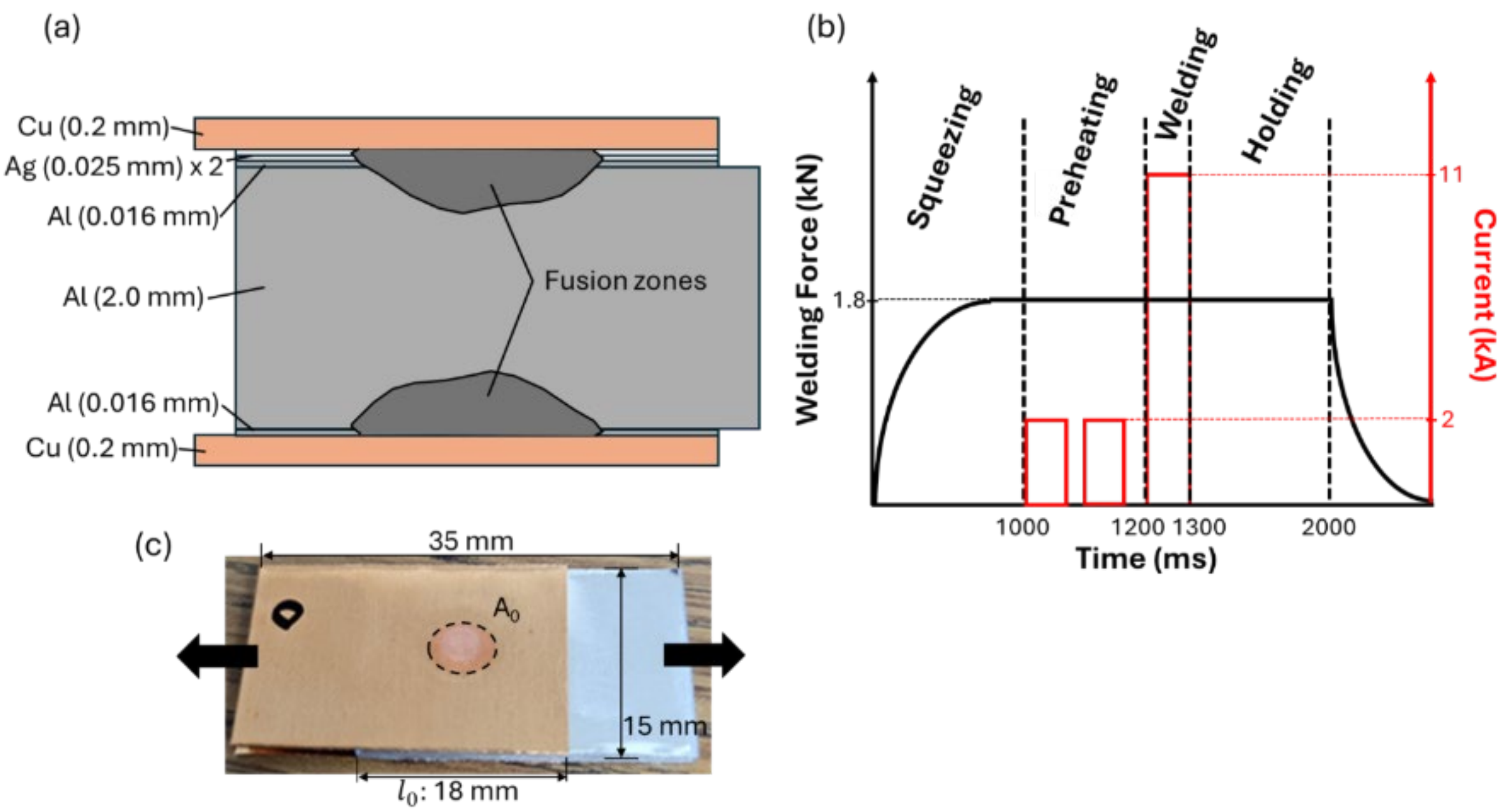

Figure 2. Materials stack up and welding procedure for Al, Ag, and Cu resistance spot welding (RSW): (a) configuration of the RSW multilayered stacks; (b) RSW procedure; and (c) configuration of tensile test sample.

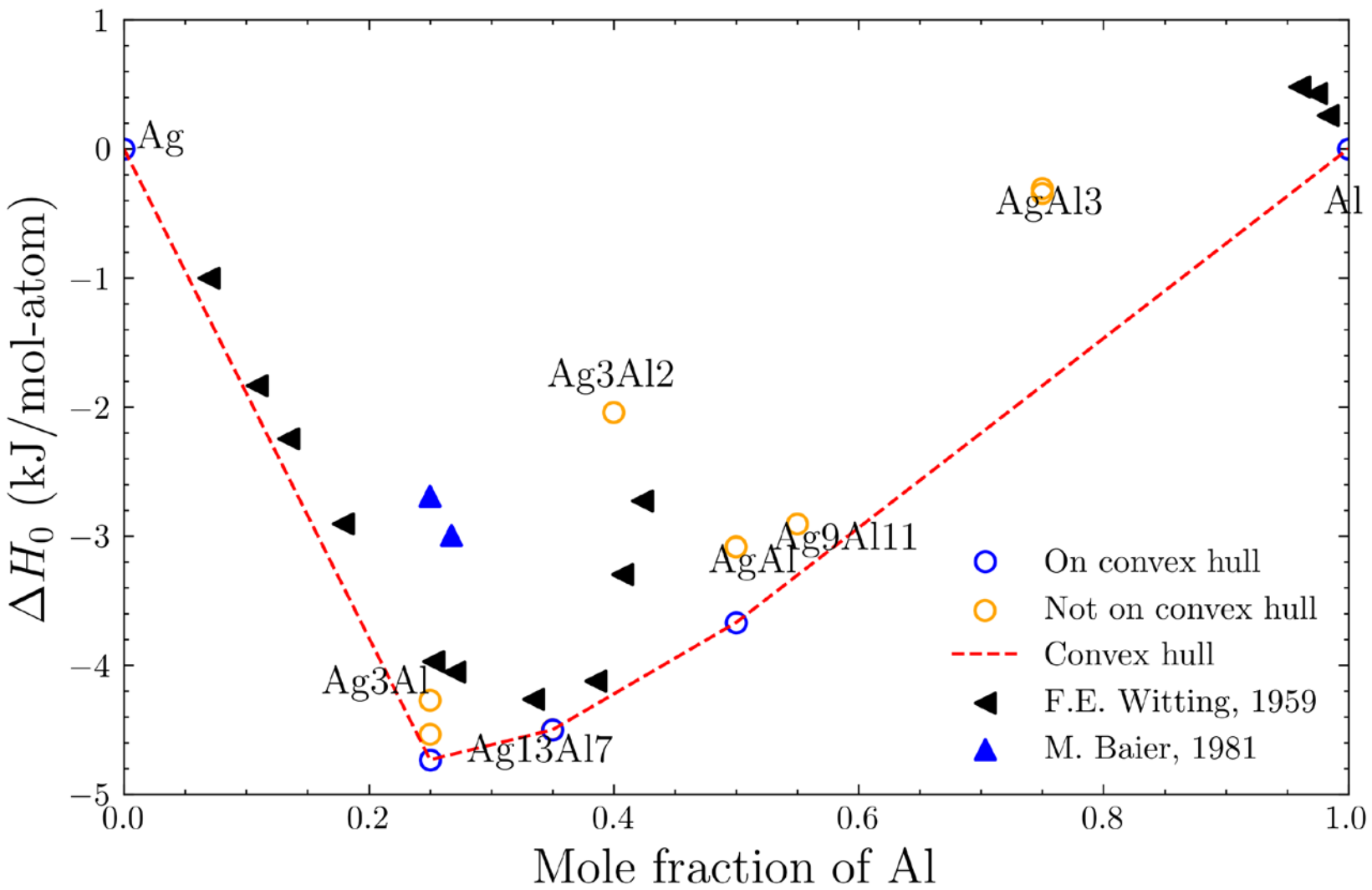

Figure 3. Convex hull of the Al-Ag system predicted by DFT formation enthalpies ($\Delta H_0$), compared with experimental formation enthalpy data from Wittig and Schilling [37] and from higher-temperature calorimetry [38]. The red dashed line denotes the convex hull by DFT, with stable compounds lying on the hull (blue circles) and metastable (or unstable) compounds positioned above the hull (orange circles).

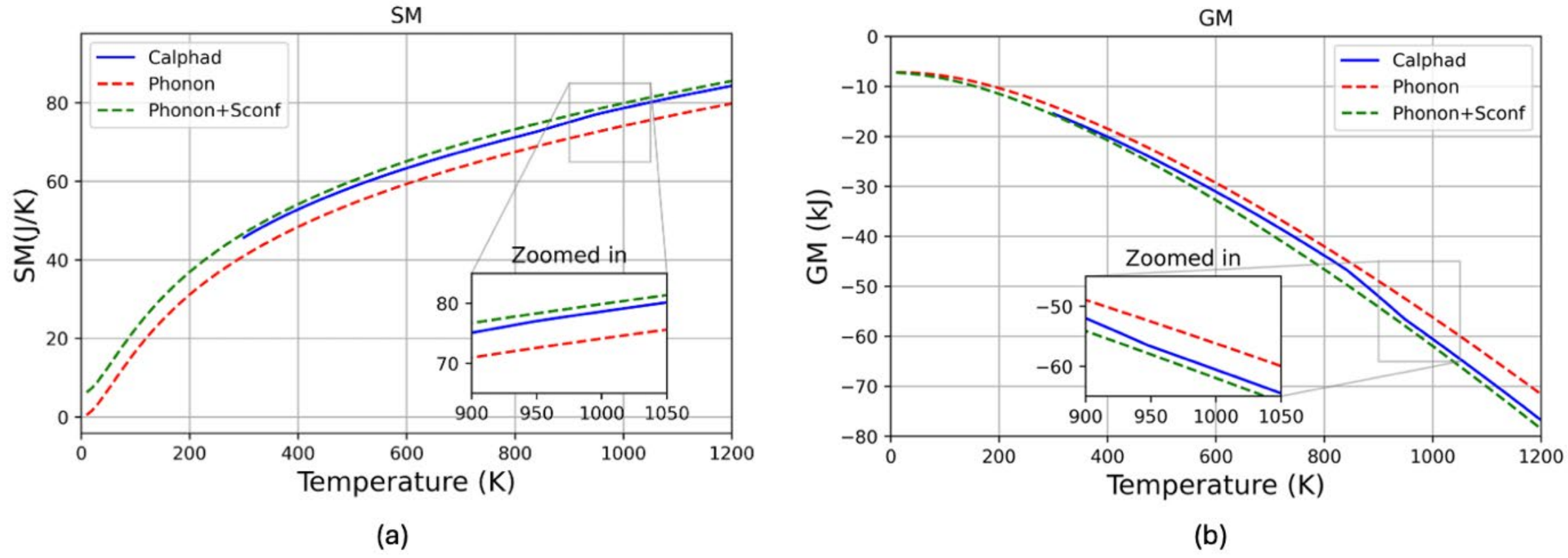

Figure 4. Comparison of temperature-dependent (a) entropy (SM) and (b) Gibbs energy (GM) of AgAl (BCC) predicted by DFT-based phonon-only, phonon + configurational entropy, and CALPHAD data [20].

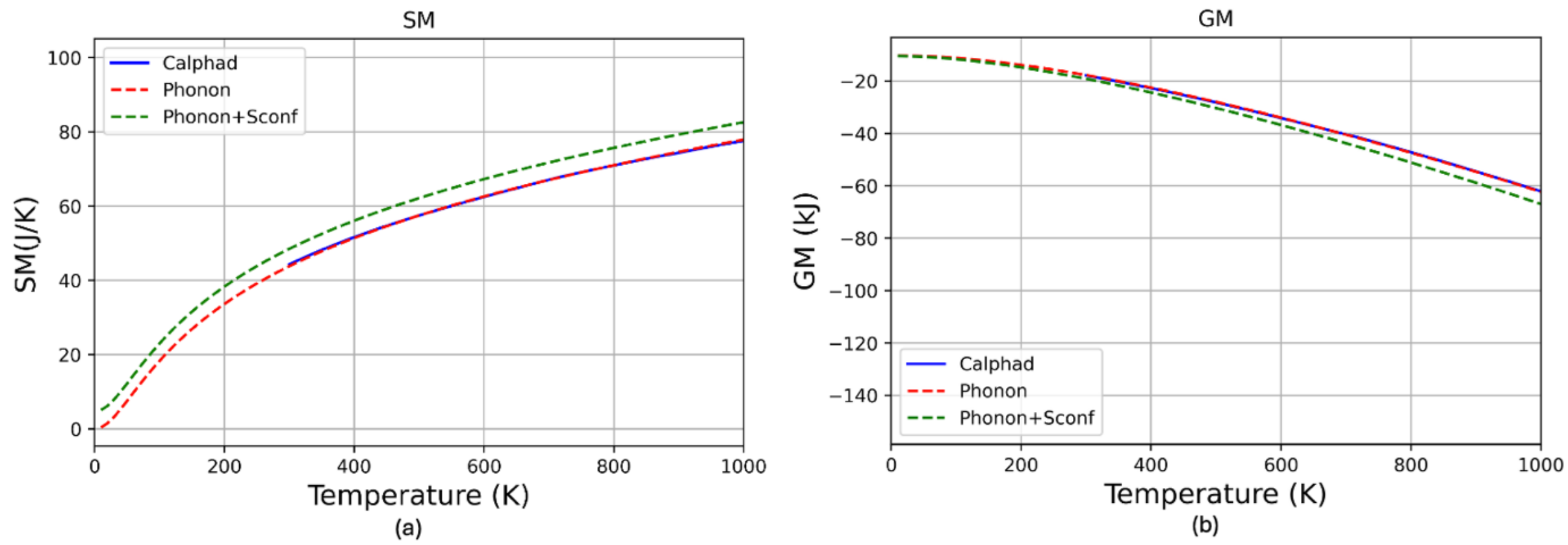

Figure 5. Comparison of temperature-dependent (a) entropy (SM) and (b) Gibbs energy (GM) of Ag3Al (HCP) predicted by DFT-based phonon-only, phonon + configurational entropy, and CALPHAD data [20].

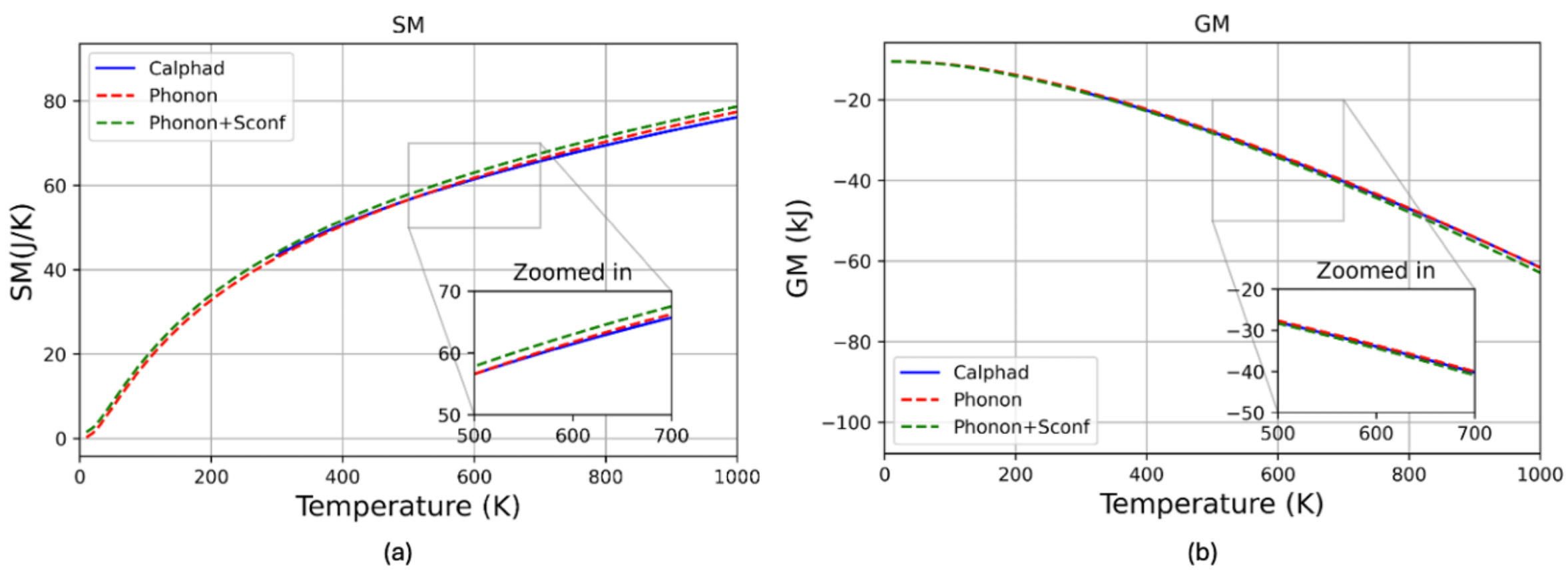

Figure 6. Comparison of temperature-dependent (a) entropy (SM) and (b) Gibbs energy (GM) of $Ag_{13}Al_7$ (A13) predicted by DFT-based phonon-only, phonon + configurational entropy, and the CALPHAD assessment of the Al-Ag system [20].

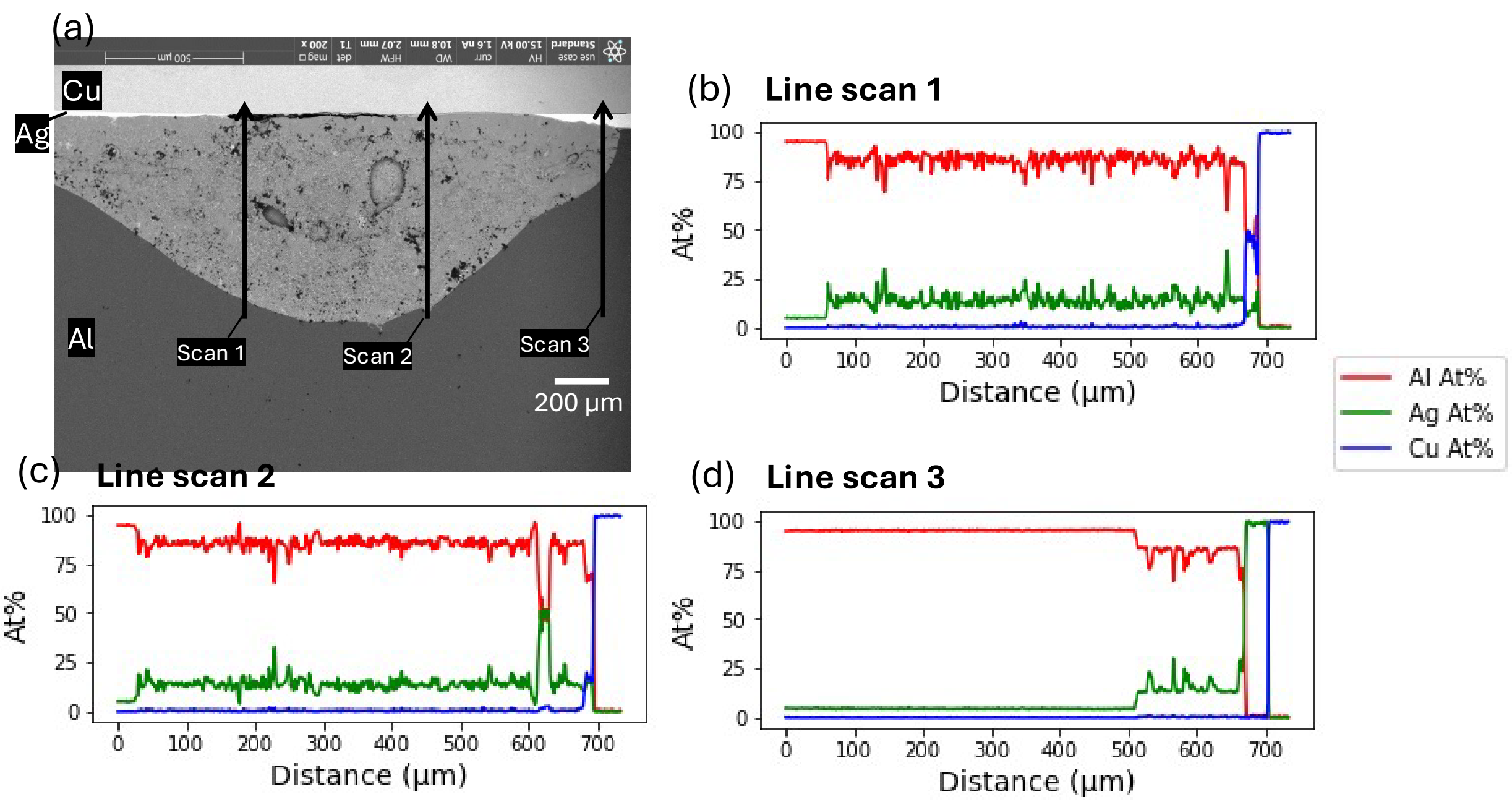

Figure 7. The SEM/EDS scanning results of the Al-Ag-Cu resistance spot weld sample: (a) SEM image with EDS line scanning regions; and (b-d) EDS line scanning results of scan 1, scan 2, scan 3, respectively.

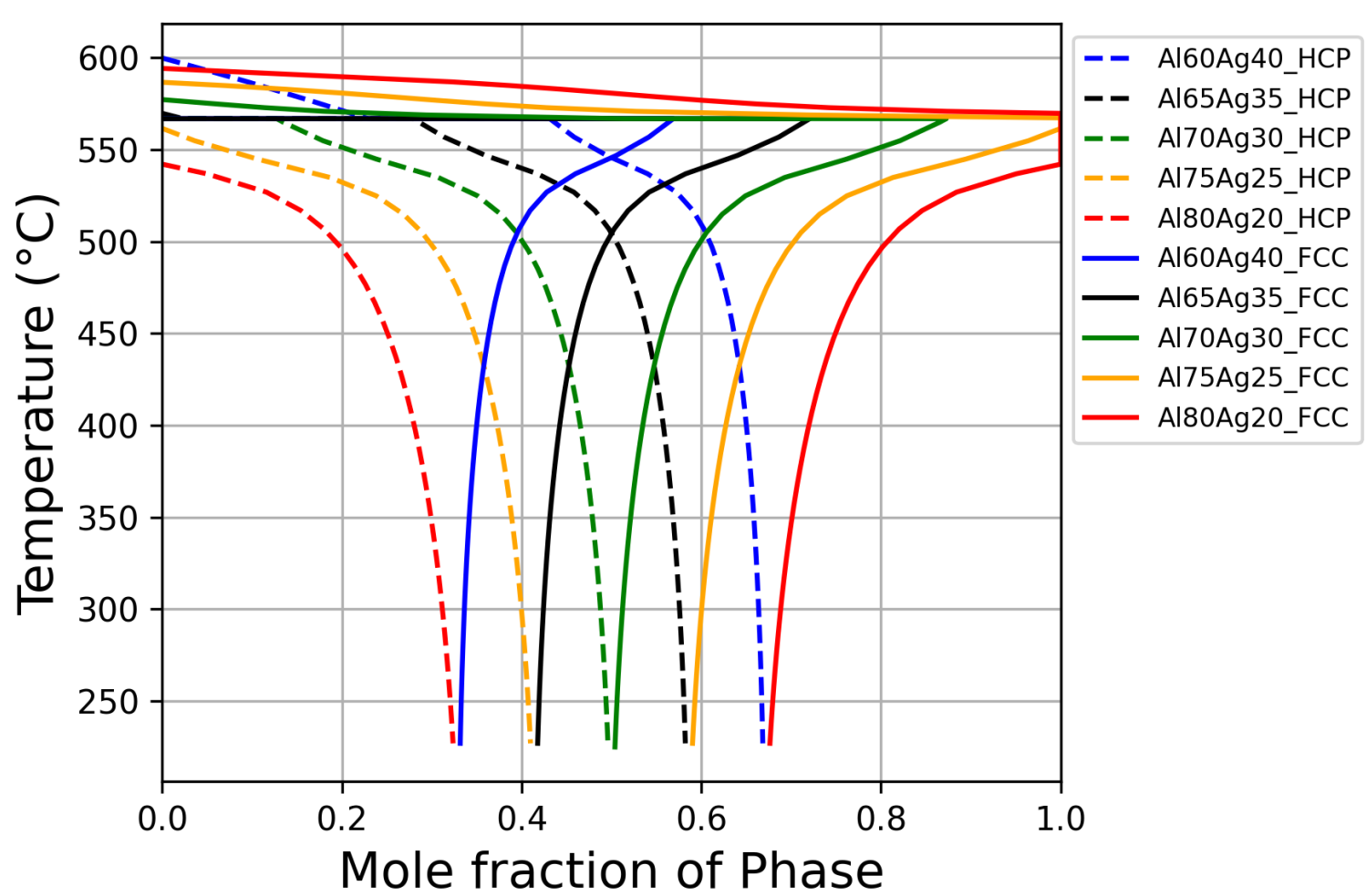

Figure 8. Calculated mole fractions of FCC and HCP phases as a function of temperature during equilibrium solidification for various Al-Ag compositions. Solid lines represent the FCC phase, while dashed lines denote the HCP phase.

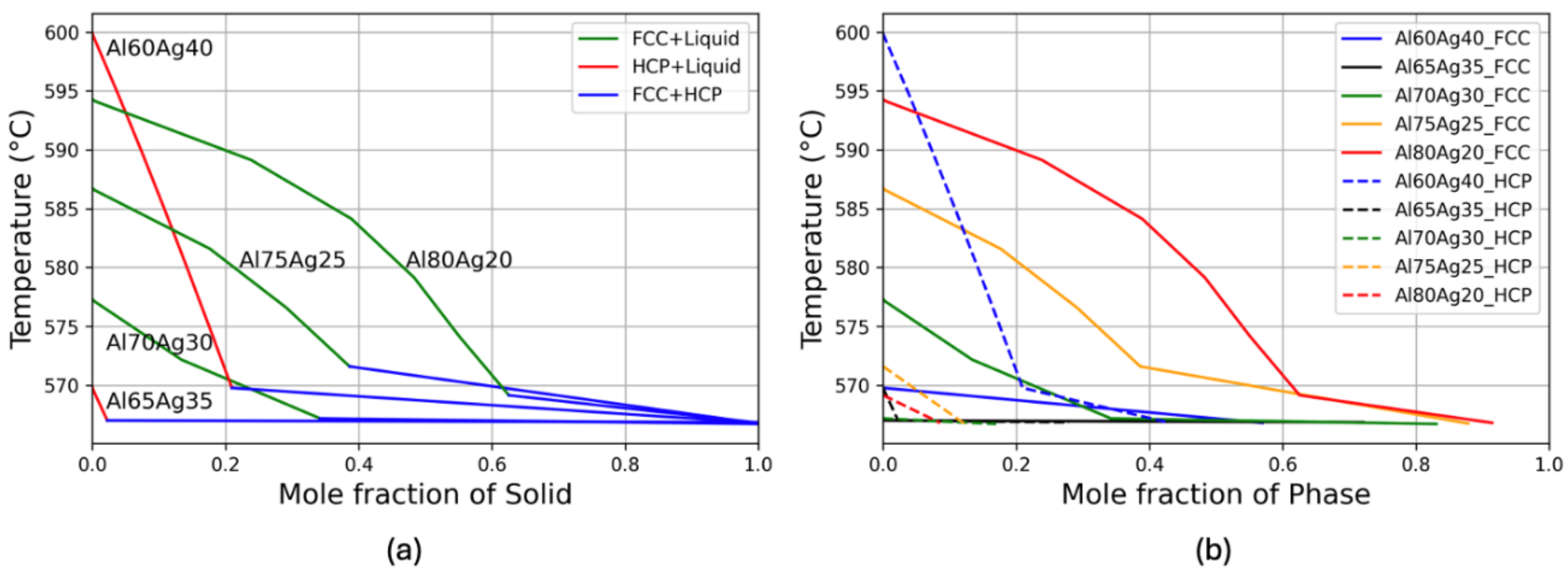

Figure 9. Scheil and Hybrid-Scheil solidification results for the five compositions ($Al_{60}Ag_{40}$, $Al_{65}Ag_{35}$, $Al_{70}Ag_{30}$, $Al_{75}Ag_{25}$, and $Al_{80}Ag_{20}$) in the Al-Ag system. (a) mole fraction of solid; (b) mole fractions of the FCC and HCP phases. Conventional Scheil and hybrid-Scheil (with solid-state back diffusion of the fast-diffusing species) results are distinguished in the legend.

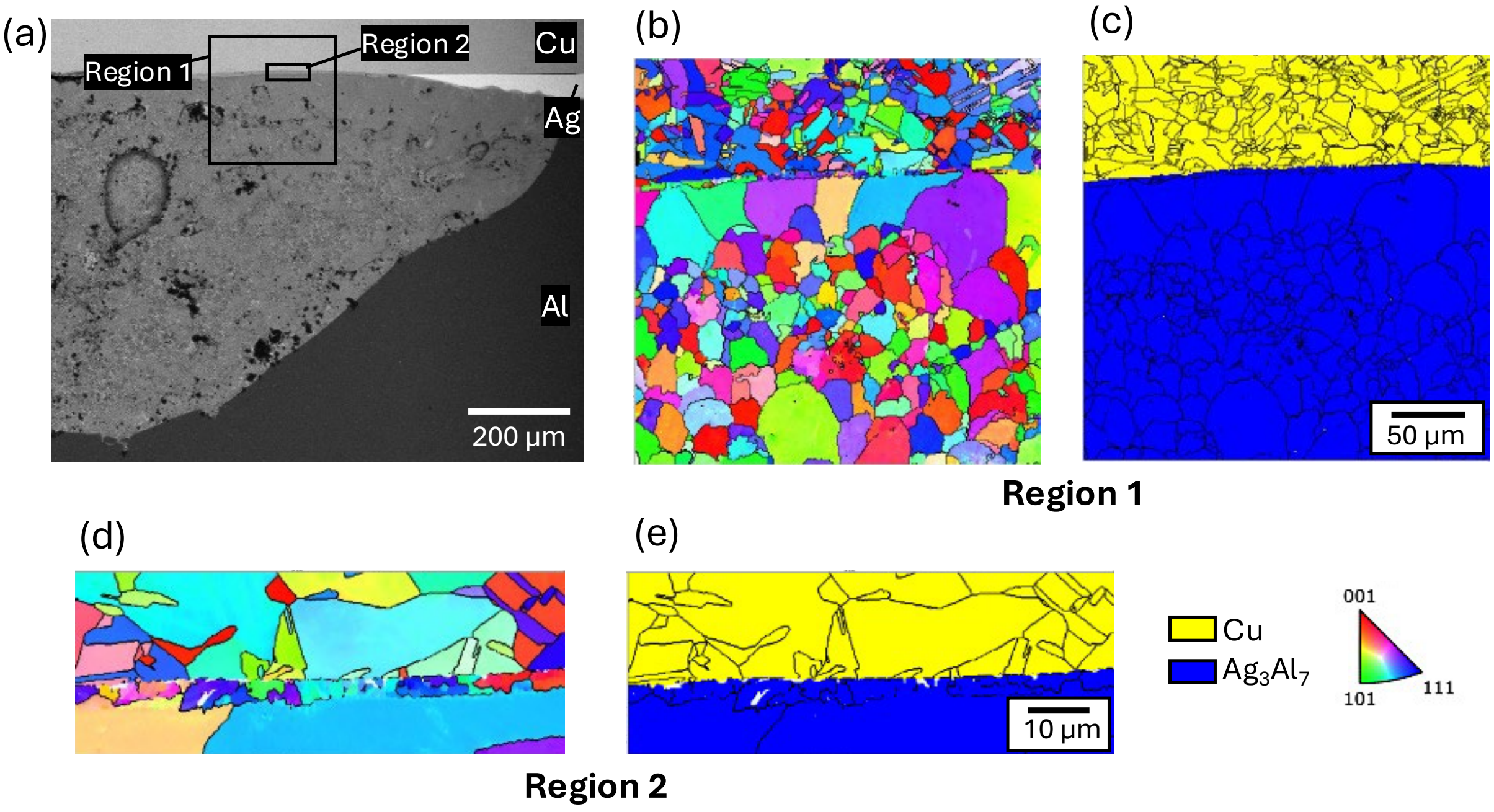

Figure 10. The EBSD scanning results of Al-Ag-Cu resistance spot weld sample: (a) SEM image with EBSD scanning regions; (b-c) inverse pole figure and phase mapping of region 1, respectively; and (d-e) inverse pole figure and phase mapping of region 2, respectively.

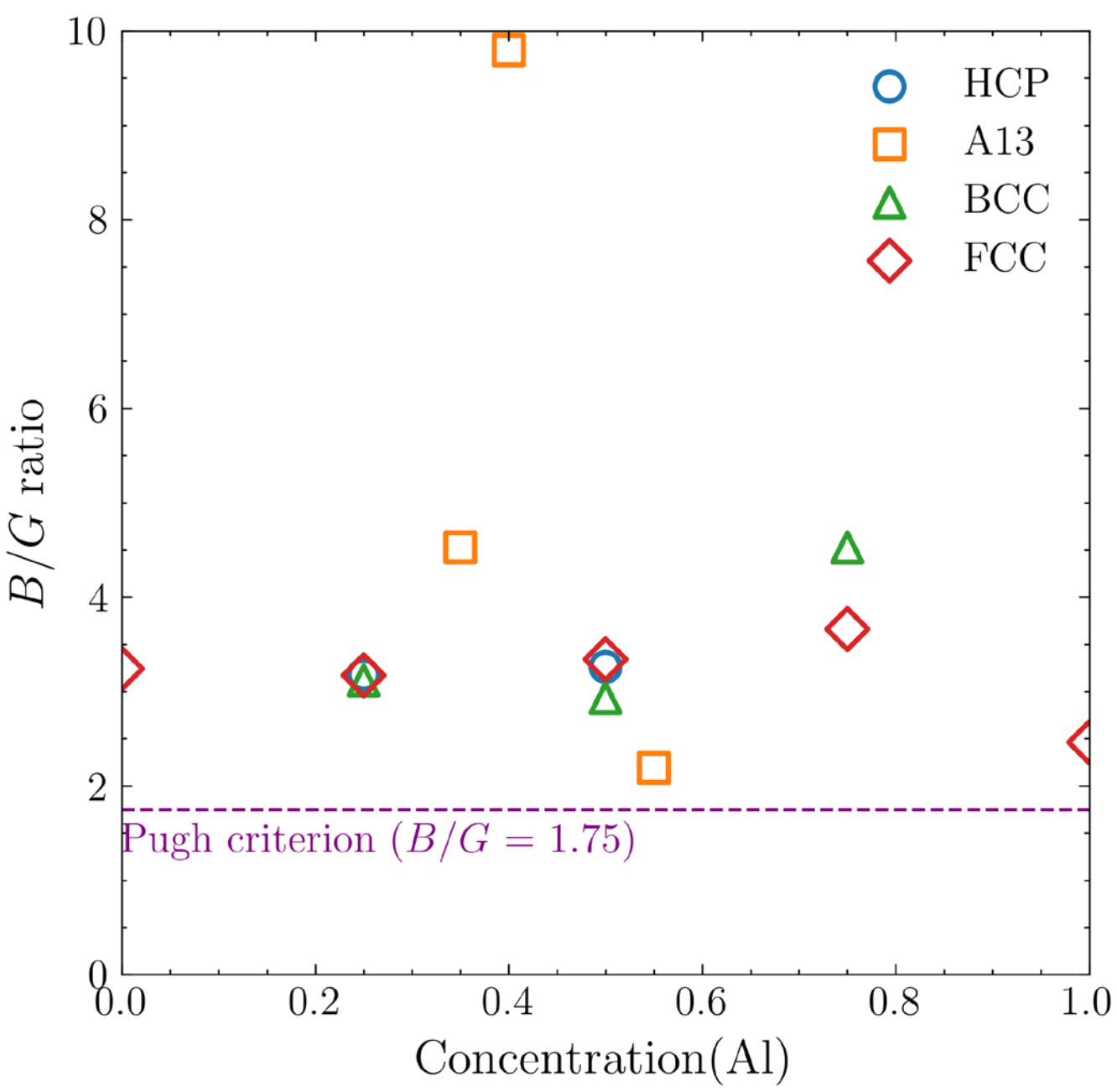

Figure 11. Calculated B/G ratios of the Al-Ag phases and solid solutions as a function of Al concentration, from the present DFT calculations; Pugh's criterion of 1.75 is a rough value to separate the ductile and brittle materials [40].

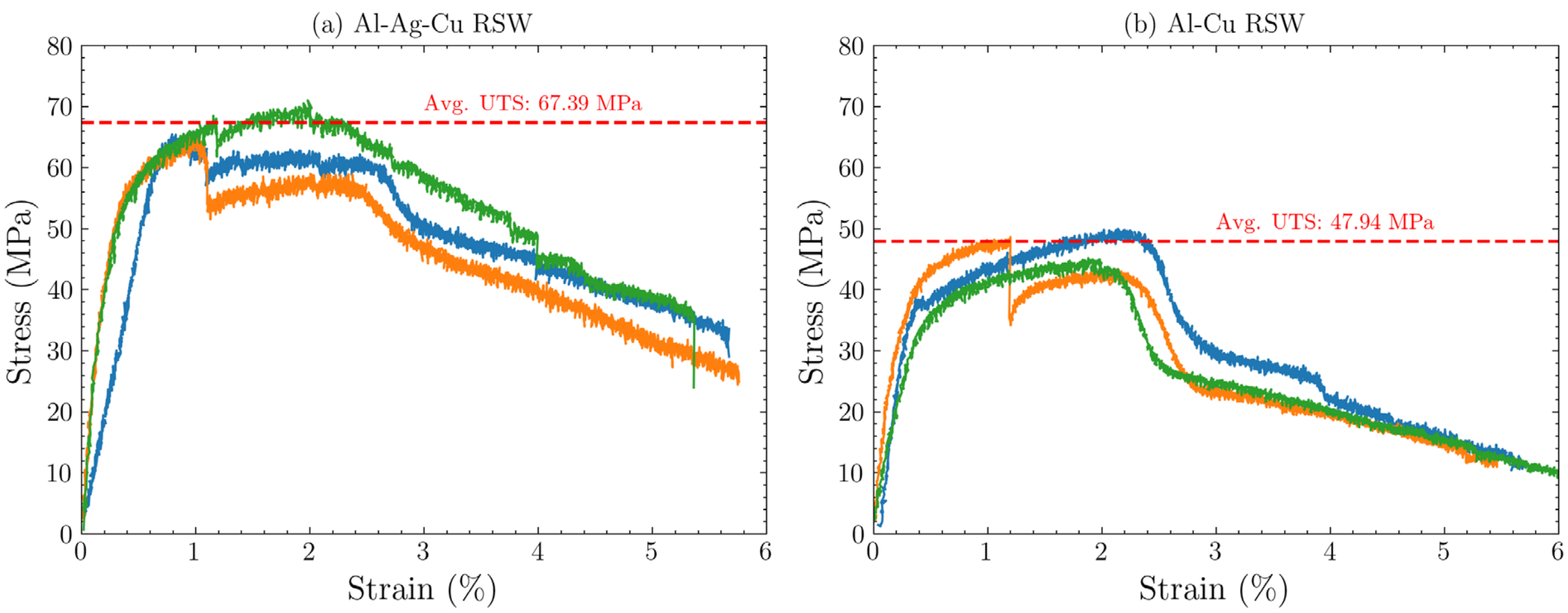

Figure 12. Nominal joint stress versus normalized crosshead displacement for resistance spot welded (RSW) joints: (a) Al-Ag-Cu system and (b) Al-Cu system. The dashed red lines mark the average peak nominal strength, 67.4 MPa for the Al-Ag-Cu joints and 47.9 MPa for the Al-Cu joints (mean of three tests).

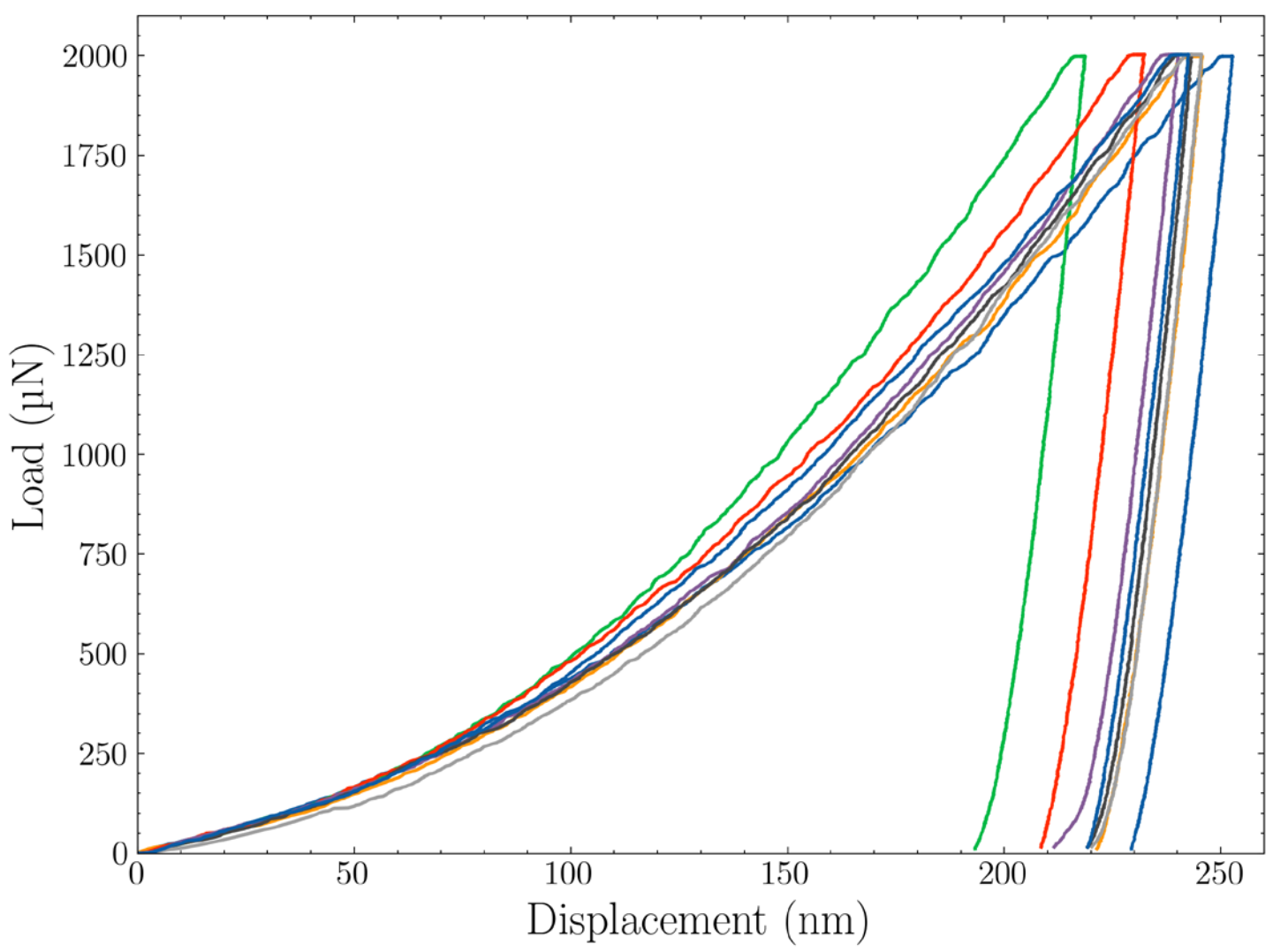

Figure 13. Nanoindentation of the Al-Ag-Cu fusion zone: (a) indentation locations within the weld zone; (b) representative load-displacement curves; and (c) indentation moduli from the eight indents, averaging 82.84 GPa.

**Tables**

Table 1. Calculated formation enthalpy (ΔH0) and mechanical properties (bulk modulus B and shear modulus G) of Al-Ag phases obtained from DFT. B and G are Voigt-Reuss-Hill averages; the Young's modulus E and Poisson's ratio ν are derived from them, and B/G is used as an empirical ductility indicator. All elastic moduli are ground-state (0 K) values. The listed space groups refer to the SQS supercells, not to the parent FCC, HCP, or BCC lattices, and 'Dilute' denotes phases modeled in the dilute-solution limit.

| **Mater.** | **Phase** | **Structure** | **Space group** | **x(Al)** | **$\Delta H_0$ (kJ mol$^{-1}$ atom)** | ***B* (GPa)** | ***G* (GPa)** | ***B/G*** | ***E* (GPa)** | **ν** |
|---|---|---|---|---|---|---|---|---|---|---|
| Ag | FCC | Endmember | $Fm\bar{3}m$ | 0 | 0 | 88.2 | 27.2 | 3.2 | 74.0 | 0.36 |
| $Ag_3Al$ | FCC | SQS | Cm | 0.25 | -4.271 | 96.1 | 26.2 | 3.7 | 72.1 | 0.38 |
| $Ag_3Al$ | HCP | SQS | P1 | 0.25 | -4.732 | 92.9 | 29.2 | 3.2 | 79.3 | 0.36 |
| $Ag_3Al$ | BCC | SQS | Cm | 0.25 | -4.535 | 93.7 | 30.1 | 3.1 | 81.6 | 0.35 |
| $Ag_{13}Al_7$ | A13 | Dilute | $P4_132$ | 0.35 | -4.500 | 92.7 | 20.4 | 4.5 | 57.0 | 0.40 |
| $Ag_3Al_2$ | A13 | Dilute | $P4_132$ | 0.4 | -2.042 | 89.6 | 9.1 | 9.8 | 26.4 | 0.45 |
| AgAl | FCC | SQS | P1 | 0.5 | -3.082 | 89.5 | 26.8 | 3.3 | 73.1 | 0.36 |
| AgAl | HCP | SQS | P1 | 0.5 | -3.085 | 86.0 | 26.4 | 3.3 | 71.8 | 0.36 |
| AgAl | BCC | SQS | P1 | 0.5 | -3.669 | 89.1 | 30.4 | 2.9 | 81.9 | 0.35 |
| $Ag_9Al_{11}$ | A13 | Dilute | $P4_132$ | 0.55 | -2.906 | 83.7 | 38.2 | 2.2 | 99.5 | 0.30 |
| $AgAl_3$ | FCC | SQS | Cm | 0.75 | -0.345 | 83.2 | 26.3 | 3.2 | 71.4 | 0.36 |
| $AgAl_3$ | BCC | SQS | Cm | 0.75 | -0.308 | 82.4 | 18.2 | 4.5 | 50.9 | 0.40 |
| Al | FCC | Endmember | $Fm\bar{3}m$ | 1 | 0 | 78.1 | 31.7 | 2.5 | 83.8 | 0.32 |